\documentclass[%
reprint,
showpacs,
nofootinbib,
amsmath,amssymb,
aps,prl
]{revtex4-1}

\usepackage{graphicx}
\usepackage{dcolumn}
\usepackage{bm}

\begin{document}

\preprint{APS/123-QED}

\title{Reduced-order modeling with artificial neurons for gravitational-wave inference}

\author{Alvin J. K. Chua}
\email{alvin.j.chua@jpl.nasa.gov}
\author{Chad R. Galley}
\email{chad.r.galley@jpl.nasa.gov}
\author{Michele Vallisneri}
\email{michele.vallisneri@jpl.nasa.gov}
\affiliation{Jet Propulsion Laboratory, California Institute of Technology, 4800 Oak Grove Drive, Pasadena, CA 91109, U.S.A.}

\date{\today}

\begin{abstract}
Gravitational-wave data analysis is rapidly absorbing techniques from deep learning, with a focus on convolutional networks and related methods that treat noisy time series as images. We pursue an alternative approach, in which waveforms are first represented as weighted sums over reduced bases (reduced-order modeling); we then train artificial neural networks to map gravitational-wave source parameters into basis coefficients. Statistical inference proceeds directly in coefficient space, where it is theoretically straightforward and computationally efficient. The neural networks also provide analytic waveform derivatives, which are useful for gradient-based sampling schemes. We demonstrate fast and accurate coefficient interpolation for the case of a four-dimensional binary-inspiral waveform family, and discuss promising applications of our framework in parameter estimation.
\end{abstract}


\maketitle

\paragraph{Introduction.}
\label{sec:introduction}

Statistical inference is the eyepiece of gravitational-wave (GW) observatories: it maps the noise-dominated detector output into probabilistic assessments of candidate significance, and into posterior probability densities for the physical parameters of confirmed detections \cite{schutz2012gravitational}. The mathematical setup of GW data analysis is simple, with most of its salient features apparent in the classical time-series likelihood
\begin{equation}
\label{eq:like}
\mathcal{L}(\theta):=p(x|\theta) \propto
\exp \left\{ -\tfrac{1}{2} \left\langle x-h(\theta)|x-h(\theta)\right\rangle \right\}.
\end{equation}
Here $x$ is the detector output, $h$ is the modeled detector response to an incoming GW with source parameters $\theta$, and $\langle \cdot | \cdot \rangle$ is the (complex) noise-weighted inner product
\begin{equation}
\label{eq:inner}
\langle a(t)|b(t)\rangle:=4\int_0^\infty\mathrm{d}f\, \frac{\tilde{a}^*(f)\tilde{b}(f)}{S_n(f)},
\end{equation}
with tildes denoting Fourier transforms and $S_n$ the one-sided power spectral density of detector noise $n$ \cite{jaranowski2009analysis}. Eq. \eqref{eq:like} essentially describes $n$ as Gaussian and additive, with the sampling distribution $p(n) \propto \exp \{ -\langle n|n \rangle^2/2 \}$.

The challenges in using \eqref{eq:like} for detection and parameter estimation are entirely practical. Detector physics poses issues such as noise characterization and response calibration; these are addressed by validating sample statistics with background studies, and including detector parameters during inference (e.g., \cite{2016PhRvD..93l2003A,2016PhRvL.116x1102A}). On the astrophysics side, the relativistic nature of many sources makes them computationally expensive to model, which hinders the bulk generation of accurate waveform templates for use in data-analysis algorithms. Furthermore, the nonlinear parameter dependence of such waveforms introduces complex features into the likelihood hypersurface (except at very high signal-to-noise ratios), making it difficult to map out with deterministic or stochastic methods.

In this Letter, we propose a general framework that combines order-reduction and machine-learning techniques to tackle these last two problems in unison, and to connect GW source modeling and data analysis in a natural and integrated manner. Our approach involves the construction of a fully representative reduced basis for the signal space of a GW model \cite{PhysRevLett.106.221102}, and the fitting of a deep neural network to this parametrized manifold. The resultant function over parameter space is an analytic waveform in reduced representation; this enables the efficient generation of signal templates, and further allows the casting of the likelihood \eqref{eq:like} in an equivalent reduced form. Also derived during the training of the network are the Jacobian and higher derivatives of the function, which encode the geometry of the signal manifold and can be used in derivative-based samplers for improved exploration of the likelihood hypersurface.

As a proof of principle, we present example results for a four-parameter post-Newtonian model of the GW signal from an inspiraling black-hole binary \cite{PhysRevD.79.104023}, which demonstrate the feasibility of our approach for higher-dimensional problems. A variety of network architectures are explored to ensure the general premise is sound. The waveform error from fitted networks approaches (to within an order of magnitude) the error in reduced-order surrogate models \cite{PhysRevX.4.031006,PhysRevLett.115.121102,PhysRevD.95.104023,PhysRevD.95.104036,PhysRevD.96.024058,2018arXiv181207865V}, and several strategies for attaining even better accuracy are outlined. We showcase the speed and robustness of our networks on a number of derivative-based applications for parameter estimation, and discuss possible extensions of the framework.

\paragraph{Reduced-order modeling.}
\label{sec:ROM}

Order-reduction strategies \cite{schilders2008} are employed in GW source modeling and data analysis to represent waveform observables as linear combinations of reduced-basis vectors \cite{PhysRevLett.106.221102,PhysRevLett.113.021101}. Unlike standard transforms, the reduced-order modeling (ROM) approach is model-specific and involves building a finite, optimally compact basis whose span is essentially the full model space (to machine precision). This reduced basis is prepared offline (i.e., in advance of its use in data analysis) by means of a greedy algorithm \cite{temlyakov_2008} that distills a large set of training templates into a smaller orthonormal set of vectors. Any waveform in model space can be reconstructed via projection onto the basis vectors, which are themselves linear combinations of training templates.

For a GW model $h(\theta)$ parametrized by $\theta\in\Theta\subset\mathbb{R}^s$, the ROM approach can be used to cast the signal space $\mathcal{S}:=h[\Theta]$ in terms of a $d$-dimensional reduced basis $\{e_i\}$ with $\langle e_i|e_j\rangle=\delta_{ij}$,\footnote{The reduced basis is orthonormal with respect to the specified noise power spectral density $S_n$ in \eqref{eq:like}; for a different $S_n$ such that $\langle e_i|e_j\rangle=N_{ij}$, pre-multiplying all coefficient vectors by the whitening matrix $W$ (where $W^\dagger W=[N_{ij}]$ is obtained through, e.g., Cholesky decomposition) restores orthonormality.} such that $\mathcal{S}$ is isomorphic to an $s$-dimensional manifold embedded in $\mathbb{C}^d$. Through projection of the signal templates onto $\{e_i\}$, we may write
\begin{equation}\label{eq:projtemplate}
h(\theta)=\sum_i\langle h(\theta)|e_i\rangle e_i:=\sum_i\alpha_i(\theta)e_i\equiv\alpha(\theta),
\end{equation}
with $\alpha\in\mathbb{C}^d$. The signal-to-noise ratio (SNR) for a template is given by $\rho:=\langle h|h\rangle^{1/2}=(\alpha^\dagger\alpha)^{1/2}:=|\alpha|$. It is convenient to work with normalized templates such that $|\alpha|=1$; for some ``true'' signal $h(\theta_*)$ with arbitrary SNR $\rho_*$, we then have $h(\theta_*)\equiv\rho_*\alpha(\theta_*)$.

The computational efficiency of reduced-order surrogates \cite{PhysRevX.4.031006,PhysRevLett.115.121102,PhysRevD.95.104023,PhysRevD.95.104036,PhysRevD.96.024058} stems from the dimensionality $d$ of the basis $\{e_i\}$ being small compared to the typical size $r$ of the standard time-series representation $h$; the linearity of $\{e_i\}$ with respect to the inner product \eqref{eq:inner} can also be exploited to accelerate likelihood evaluations, through the method of reduced-order quadratures \cite{PhysRevD.87.124005,PhysRevLett.114.071104}. However, all of these benefits rely on being able to obtain (for arbitrary source parameters $\theta$) the basis projection coefficients $\alpha(\theta)$ without computing the full waveform $h(\theta)$ itself. Present applications make use of empirical interpolation \cite{barrault2004empirical}, which requires $h$ to be evaluated (or approximated) only at a set of $d$ time/frequency nodes that are uniquely defined by the reduced basis. Direct interpolation of $\alpha$ across parameter space remains challenging, except in low-dimensional ($s\lesssim3$) problems.

\paragraph{Artificial neural networks.}
\label{sec:ANN}

The deep-learning paradigm encompasses a family of machine-learning techniques that automatically discover and process ``features'' in data, without the need for task-specific algorithms \cite{Goodfellow-et-al-2016}. Most methods in deep learning are based on variants of artificial neural networks (ANNs)\,---\,biologically inspired computational objects comprising multiple layers of nonlinear processing nodes between the input to the network and its output. Within GW data analysis, ANNs trained under supervision (i.e., using data structured in the form of input--output pairs \cite{russell2010artificial}) are potential ``black-box'' alternatives to more statistically principled methods based on Eq.\,\eqref{eq:like}, and they have recently been shown to achieve promising performance in detection and classification tasks \cite{GebKilParHarSch17,PhysRevD.97.044039,GEORGE201864,0264-9381-35-9-095016,PhysRevLett.120.141103,PhysRevD.97.101501,2018arXiv180709787R}.

ANNs are also well known to be universal approximators \cite{Cybenko1989} for continuous functions: given an appropriate learning strategy and properly structured training data, a network of sufficient depth can interpolate between training examples to a high level of accuracy. In this Letter, we demonstrate that ANNs are in principle suited to the high-dimensional interpolation problem posed by the ROM projection coefficients. We design networks that take $\theta$ as their input, and output $\hat{\alpha}(\theta)$ (where the overhat is used here and henceforth to denote an interpolated estimate). These are fitted to a large set of training pairs $\{\theta_n,\alpha(\theta_n)\}$ chosen to adequately represent the domain of interest $\Theta$; goodness of fit is then evaluated on a small test set of examples that are held out from the training stage. The result is a function $\hat{\alpha}(\theta)\approx\alpha(\theta)$ that is both fully analytic and computationally efficient, as it is composed of (many) closed-form array operations.

Convolutional-type network architectures \cite{LeCun:1998:CNI:303568.303704} have become the dominant model in deep learning, due to high-profile successes in applications such as computer vision \cite{sebe2005machine} and natural language processing \cite{Goldberg:2016:PNN:3176748.3176757}. They leverage spatial correlations in high-resolution data to lower the number of free parameters in the network, which in turn reduces overfitting. However, our investigations with deconvolutional neural networks \cite{deconv} indicate that they are less suited to the ROM interpolation problem; here training data is abundant, and underfitting is more of an issue. This is because ROM coefficient data has already been pared down to its principal features, and in its present vector form cannot be organized spatially (although this might be possible with tensor decompositions/networks \cite{2014arXiv1403.2048C}). The low resolution ($d\sim10^2$) of the data makes it more amenable to the ``fully connected'' layers in a multilayer perceptron \cite{haykin1999neural}, which are individually faster than convolutional layers at low width, and hence can be stacked to great depth for increased network capacity.

In our multilayer perceptron networks, the input layer $\theta$ and output layer $\hat{\alpha}$ are connected by a sequence of ``hidden'' layers, each parametrized by a matrix of weights $w$ and a vector of biases $b$. The $\ell$-th hidden layer takes an input $a_{\ell-1}$ from the previous layer, and outputs to the next layer the value $a_\ell=a(w_\ell a_{\ell-1}+b_\ell)$ of a closed-form, nonlinear ``activation'' function $a$. Values of the weights and biases that minimize a suitably defined loss of fidelity $L$ in the output are learnt during the training stage, where the loss gradient $\partial L/\partial(w,b)$ is obtained through a backpropagation algorithm \cite{Rumelhart:1986:LIR:104279.104293} and used iteratively in gradient descent optimization. ANNs can also be used to compute the analytic derivatives $\partial^n\hat{\alpha}/\partial\theta^n$ as functions of $(w,b)$ (for $n-1$ up to the differentiability class of $a$); these converge to the target derivatives $\partial^n\alpha/\partial\theta^n$ with no added computational expense as the network is trained.

\paragraph{The reduced likelihood.}
\label{sec:redlike}

We now consider the basic case where the data $x$ in \eqref{eq:like} is the sum of a single true signal $h(\theta_*)$ and the detector noise $n$. Projecting the data onto the reduced basis $\{e_i\}$ as in \eqref{eq:projtemplate} gives $x\equiv\beta+\gamma$; here the reduced-space term $\beta=\rho_*\alpha(\theta_*)+\nu$ contains the true signal template $\alpha(\theta_*)$ and a noise component $\nu\equiv\sum_i\langle n|e_i\rangle e_i$, while the orthogonal term $\gamma=x-\beta$ is the projection of $n$ onto the orthogonal complement of $S$. In this representation, the likelihood \eqref{eq:like} becomes
\begin{align}\label{eq:redlike}
\mathcal{L}(\theta)&\propto\exp{\left\{-\tfrac{1}{2}|\beta-\alpha(\theta)|^2-\tfrac{1}{2}|\gamma|^2\right\}},\nonumber\\&\propto\exp{\left\{-\tfrac{1}{2}|\beta-\alpha(\theta)|^2\right\}},
\end{align}
since the orthogonal noise $\gamma$ does not depend on $\theta$.

Eq.\,\eqref{eq:redlike} corresponds to the trivial probability model $\nu\sim\mathcal{N}(0,I_d)$ for detector noise in the reduced space $\mathbb{C}^d$, and retains the statistical properties of Eq.\,\eqref{eq:like}. Computation of the $\theta$-dependent terms $\beta^\dagger\alpha(\theta)$ and $|\alpha(\theta)|^2$ in \eqref{eq:redlike} is an $\mathcal{O}(d)$ operation, as compared to $\mathcal{O}(r\gg d)$ for $\langle x|h(\theta)\rangle$ and $\langle h(\theta)|h(\theta)\rangle$ in \eqref{eq:like} (where $r$ is the number of frequency-series samples).\footnote{The reduced likelihood \eqref{eq:redlike} shares the start-up cost of projecting the data with the reduced-order quadrature method \cite{PhysRevD.87.124005,PhysRevLett.114.071104}, but there the cost of the template--template term scales as $\mathcal{O}(d^2)$.} With the efficient generation of $\hat{\alpha}(\theta)$ provided by a neural network, and the explicit reduction of complexity $r\to d$ in the reduced likelihood \eqref{eq:redlike}, much of the computational cost associated with online likelihood evaluations can thus be shifted into the offline construction of the reduced basis and ANN interpolant. If so desired, extrinsic parameters such as amplitude or time of arrival can still be handled analytically, as is customary in the GW literature \cite{2003PhRvD..67b4016B}.

\paragraph{Example results.}
\label{sec:results}

\begin{figure}[!tbp]
\centering
\includegraphics[width=\columnwidth]{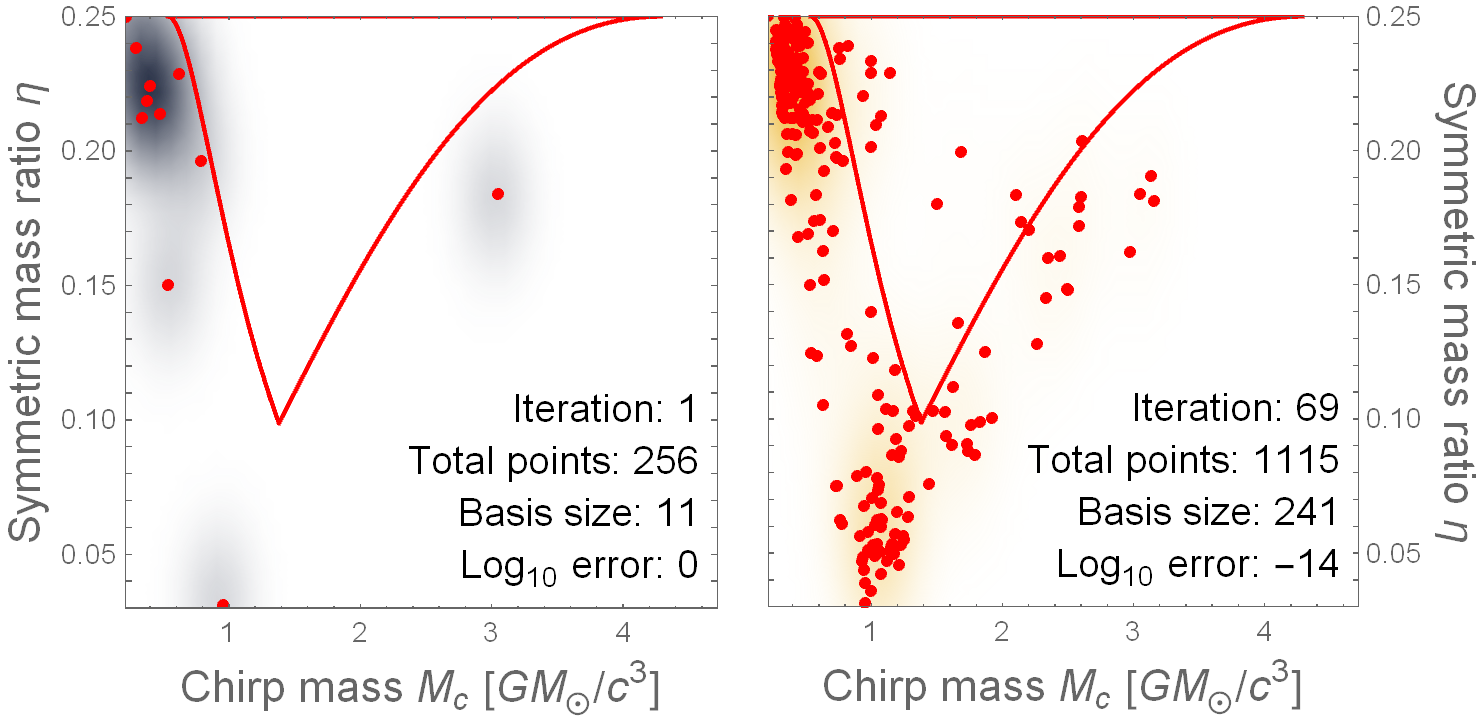}
\caption{Starting with a reduced basis built from 256 training points distributed uniformly over an extended domain, we draw up to 20 more training points from the kernel density estimate of the basis points at each iteration (red points, projected onto the $(M_c,\eta)$-plane), and construct a new basis. A final basis representation error of $\lesssim10^{-14}$ over the domain of interest (red border) is attained with only $\sim10^3$ training points, which is far more efficient than grid-based training.}
\label{fig:kde}
\end{figure}

In this work, we apply our approach to a 2.5PN TaylorF2 waveform family \cite{PhysRevD.79.104023}. This analytic frequency-domain model describes the GW signal from an inspiraling black-hole binary with aligned spins, and is parametrized by the component masses $m_{1,2}$ and dimensionless spins $\chi_{1,2}$. We consider signals with $r=10^4$ frequency samples over the range $[1,10]\,\mathrm{mHz}$, and a parameter domain defined by $m_{1,2}\in[1.25,10]\times10^5M_\odot$ and $\chi_{1,2}\in[-1,1]$; these choices correspond to a subset of the high-redshift massive-black-hole binaries that will be observed by the space-based GW detector LISA \cite{2017arXiv170200786A}.

Working in the more natural mass parametrization of chirp mass $M_c$ and symmetric mass ratio $\eta$ \cite{PhysRevD.49.2658}, we construct a reduced basis for the above model, but over an extended domain with $m_{1,2}\in[0.5,15]\times10^5M_\odot$ and $\chi_{1,2}\in[-1.5,1]$. This facilitates a new high-dimensional ROM training strategy \cite{kde} that iteratively builds up a kernel density estimate for the parameter-space distribution of templates selected by the greedy algorithm; in the present model, these tend to cluster at the low-mass and negative-spin boundaries (see Fig.\,\ref{fig:kde}). A subset of the extended domain is also used in the training of the ANNs, where compensating for the added structure helps to improve accuracy in the domain of interest. For the four-dimensional model, the size of the reduced basis is $d=241$. A smaller basis is derived for the non-spinning two-dimensional submodel over the same range of masses.

Instead of interpolating $\alpha(\theta)\in\mathbb{C}^{d}\cong\mathbb{R}^{2d}$ with a single network, it is more practical to train two independent ANNs on its real and imaginary parts; this allows the layer width to be kept small, and it is straightforward to combine the two network outputs post hoc. We implement our ANNs using the standard \texttt{TensorFlow} software library \cite{tensorflow2015-whitepaper}. Our network for the four-dimensional model contains 25 hidden layers $a_\ell$, where the first five comprise $2^{\ell+2}$ nodes and the remaining layers have 256 nodes each. This choice of architecture is arrived at heuristically, with the maximal layer width determined by that of the output layer, and with 256-node layers being appended to a smaller initial network until a satisfactory level of accuracy is achieved (without overfitting).

While the ROM training set is far too sparse for interpolation purposes, it captures much of the underlying structure and might be used to inform the distribution of a larger ANN training set. However, a uniform grid of training examples over an extended domain can also yield good accuracy in the domain of interest. For the four-dimensional model, $6\times10^5$ training examples are needed to prevent overfitting, which we define as a difference of $>0.1\%$ in median accuracy on the training and test sets. To enforce this, small validation sets of 5,000 examples are randomly generated throughout the training stage, and are used to inform the regularization method of early stopping \cite{623200}. Early stopping turns out to be unnecessary for the final network and full training set, since underfitting is present instead; this is indicated by a leveling-off in performance on both the training and validation sets as the network is trained.

The ``leaky RELU'' activation function \cite{Maas2013RectifierNI} is applied on all hidden layers of the four-dimensional network, while linear activation (with $a$ being the identity) is used on the output layer. Although the leaky RELU has good training efficiency \cite{pmlr-v15-glorot11a}, it is of class $C^0$ with vanishing second derivative, such that $\partial^2\hat{\alpha}/\partial\theta^2$ also vanishes globally. This is not the case for the (slower-to-train) $\tanh$ activation function \cite{LeCun2012}, which we employ in an ANN for the two-dimensional submodel. A mini-batch \cite{Goodfellow-et-al-2016} size of $10^3$ is empirically chosen for the training stage, and the adaptive, momentum-based ADAM optimization algorithm \cite{2014arXiv1412.6980K} (but with a manually decayed initial learning rate) is used to minimize the loss function\footnote{This least-squares loss is weighted to converge from the direction of a larger norm, which helps to preserve the proportion between the independently trained real and imaginary parts.}
\begin{equation}\label{eq:loss}
L:=\frac{\left\langle|\alpha-\hat{\alpha}|^2\right\rangle}{\sqrt{\left\langle|\hat{\alpha}|^2\right\rangle}},
\end{equation}
where $\langle\cdot\rangle$ denotes the average over a mini-batch.

\begin{figure}[!tbp]
\centering
\includegraphics[width=\columnwidth]{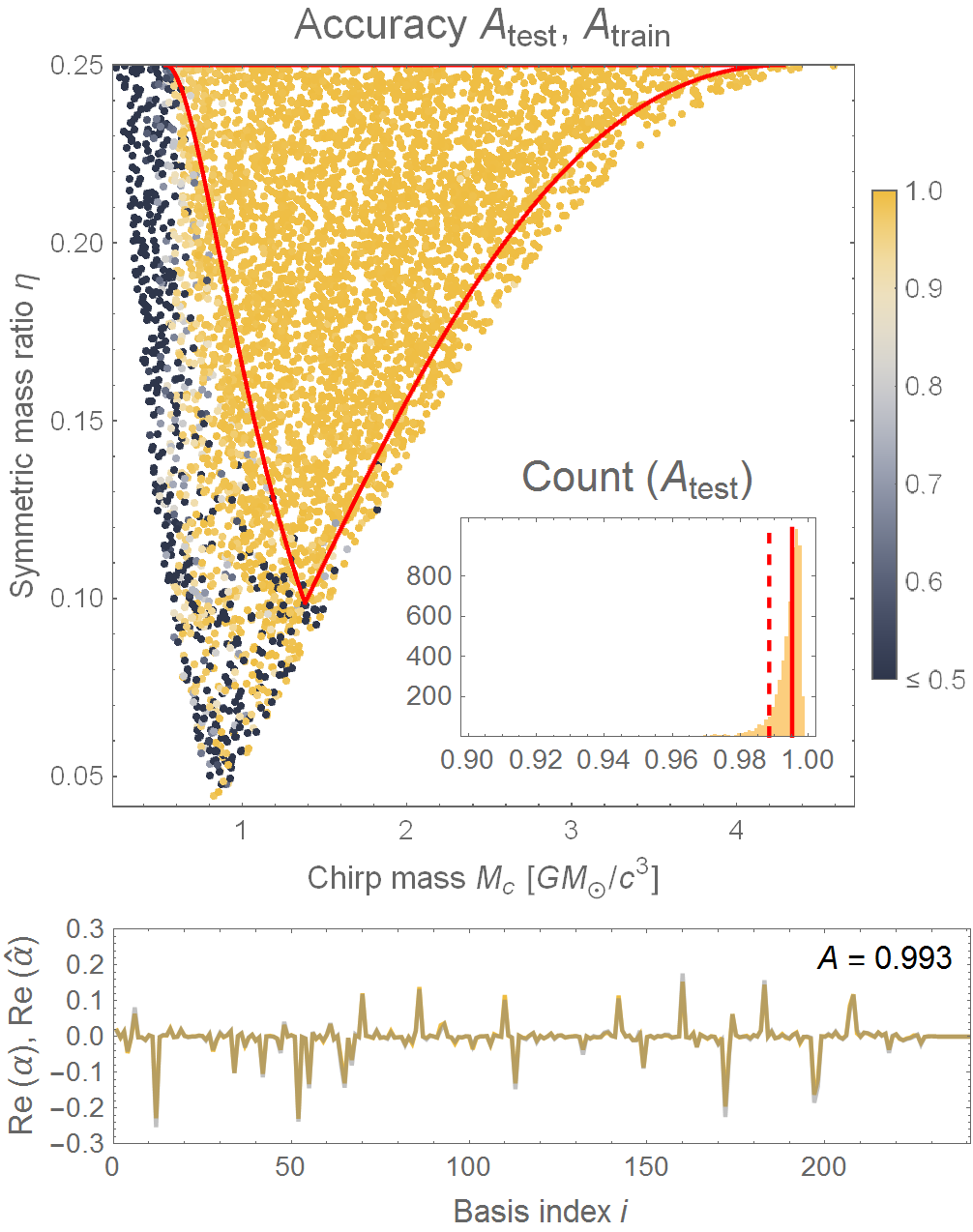}
\caption{Top: Plot of accuracy $A$ as a function of $(M_c,\eta)$ for test set (inside red border), and for 3,000 training examples with $(M_c,\eta)$ outside the domain of interest. Inset: Histogram of test-set accuracy values with tenth percentile (dashed line) and median (solid line) indicated. Bottom: Visualization of typical coefficient vectors $\mathrm{Re}(\alpha)$ (yellow) and $\mathrm{Re}(\hat{\alpha})$ (gray).}
\label{fig:accuracy}
\end{figure}

\begin{figure}[!tbp]
\centering
\includegraphics[width=\columnwidth]{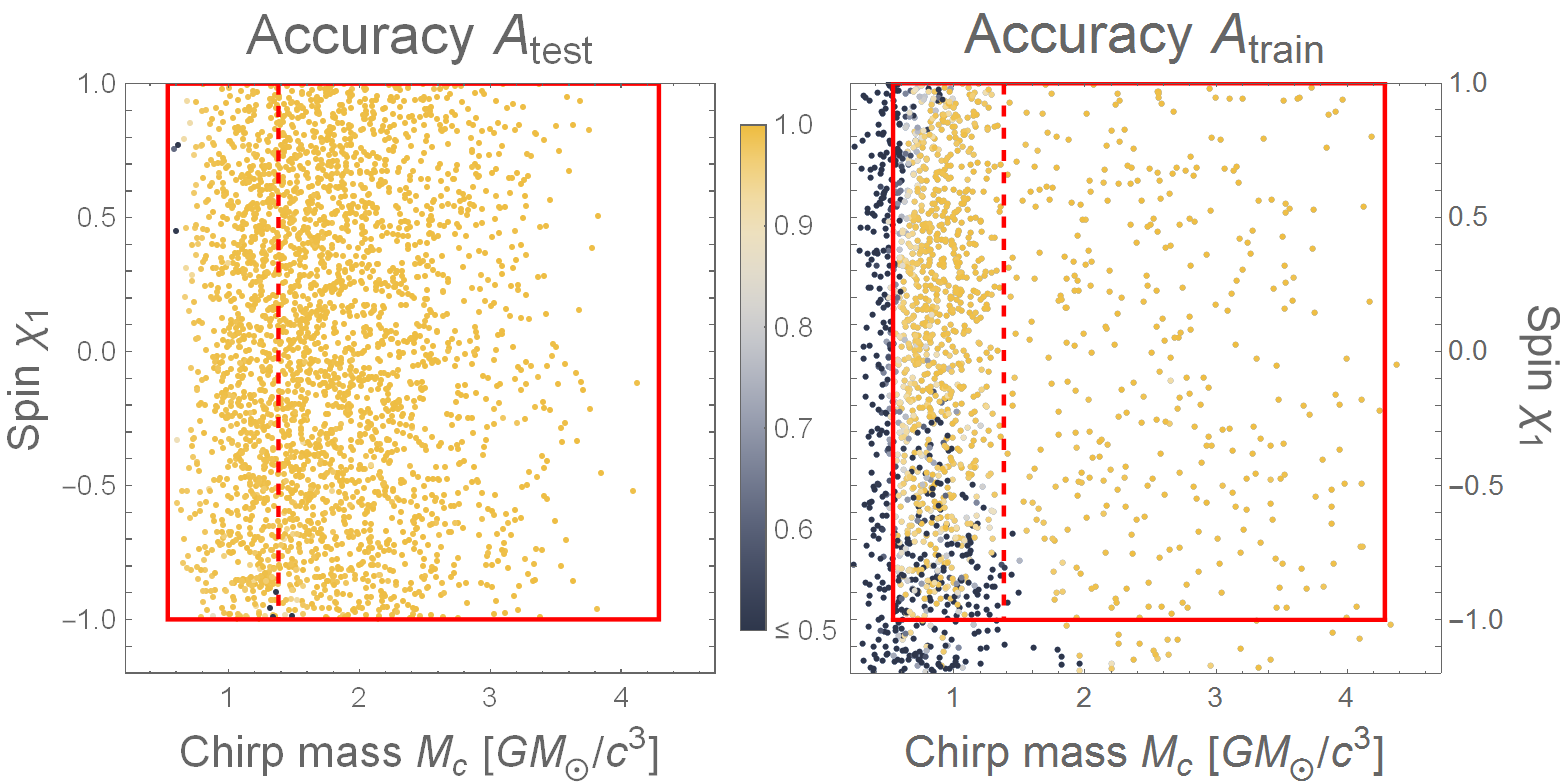}
\caption{Plot of accuracy $A$ as a function of $(M_c,\chi_1)$ for test set (left), and for the 3,000 training examples in Fig.\,\ref{fig:accuracy} (right).}
\label{fig:massspin}
\end{figure}

To quantify the accuracy of $\hat{\alpha}$ on a test set of 5,000 examples, we use the normalized inner product (or overlap) between $\alpha$ and $\hat{\alpha}$, i.e., $A:=\alpha^\dagger\hat{\alpha}/|\hat{\alpha}|$. The error $1-A$ is related to \eqref{eq:loss} for a single template by $1-A\leq L/2$, with equality in the case of an accurate norm ($|\hat{\alpha}|=1$). Results for the four-dimensional model are presented in Figs \ref{fig:accuracy} and \ref{fig:massspin}. The ANN achieves a median accuracy of 99.5\%; other high-dimensional ROM applications such as the numerical-relativity surrogates have typical median accuracies of 99.9\% \cite{PhysRevD.95.104023,PhysRevD.96.024058}. Some of the disparity can be attributed to the four-fold larger mass ratio of eight in our model. Regardless, we estimate that the error in our ANN decays exponentially with network capacity: no more than 40 layers should be enough to hit 99.9\%, which would only raise the network evaluation time from 1.5 to $2\,\mathrm{ms}$ (although the training-set size and the number of training epochs would have to be increased as well).

\paragraph{Parameter-estimation applications.}
\label{sec:applications}

\begin{figure}[!tbp]
\centering
\includegraphics[width=\columnwidth]{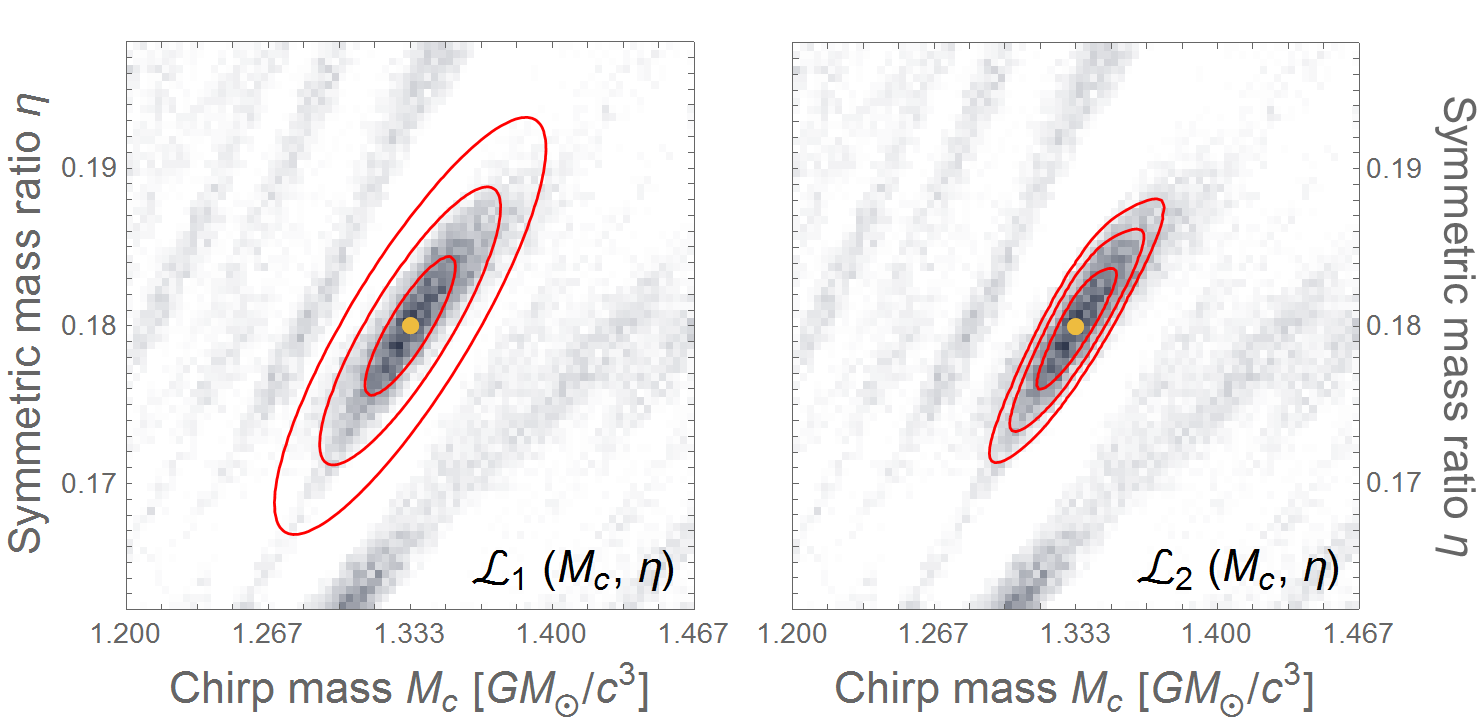}
\caption{One- to three-sigma contours (red) for $\mathcal{L}_1$ (left) and $\mathcal{L}_2$ (right) overlaid on density histogram of $10^5$ samples drawn from $\mathcal{L}$, with $\theta_\mathrm{ML}=\theta_*=(4/3,0.18)$ (yellow point).}
\label{fig:fisher}
\end{figure}

The analytic waveform derivatives from our ANNs hold intriguing possibilities for GW data analysis, as they are immune to the speed and stability issues posed by taking derivatives numerically. For the above four-dimensional model, the Jacobian $J^a_{\;i}:=\partial_i\hat{\alpha}^a$ is evaluated in $\sim 0.1\,\mathrm{s}$ (without optimization); this yields fast and accurate estimates of the Fisher information matrix $F_{ij}:=J_{ai}J^a_{\;j}$. The Fisher matrix describes the linearized (in $\hat{\alpha}$) likelihood around the maximum-likelihood estimate $\theta_\mathrm{ML}$ \cite{PhysRevD.77.042001}, i.e.,
\begin{equation}\label{eq:firstlike}
\mathcal{L}_1(\theta)\propto\exp{\left\{-\tfrac{1}{2}F_{ij}\vartheta^i\vartheta^j\right\}},
\end{equation}
where $\vartheta:=\theta-\theta_\mathrm{ML}$. Eq.\,\eqref{eq:firstlike} is inaccurate at low detection SNR, but may be improved with the Hessian $H^a_{\;ij}:=\partial_i\partial_j\hat{\alpha}^a$. The noise-free second-order likelihood is \cite{PhysRevD.77.042001,doi:10.1093/mnras/stu689}
\begin{equation}\label{eq:secondlike}
\mathcal{L}_2(\theta)\propto\mathcal{L}_1(\theta)\exp\left\{-\tfrac{1}{2}C_{ijk}\vartheta^i\vartheta^j\vartheta^k-\tfrac{1}{8}Q_{ijkl}\vartheta^i\vartheta^j\vartheta^k\vartheta^l\right\},
\end{equation}
where $C_{ijk}:=J_{ai}H^a_{\;jk}$ and $Q_{ijkl}:=H_{aij}H^a_{\; kl}$. In Fig.\,\ref{fig:fisher}, we compare the probability contours of $\mathcal{L}_1$ and $\mathcal{L}_2$ to the stochastically sampled $\mathcal{L}$, for a low-SNR injected signal $\beta=2\hat{\alpha}(\theta_*)$
in the two-dimensional $(M_c, \eta)$-submodel.

\begin{figure}[!tbp]
\centering
\includegraphics[width=\columnwidth]{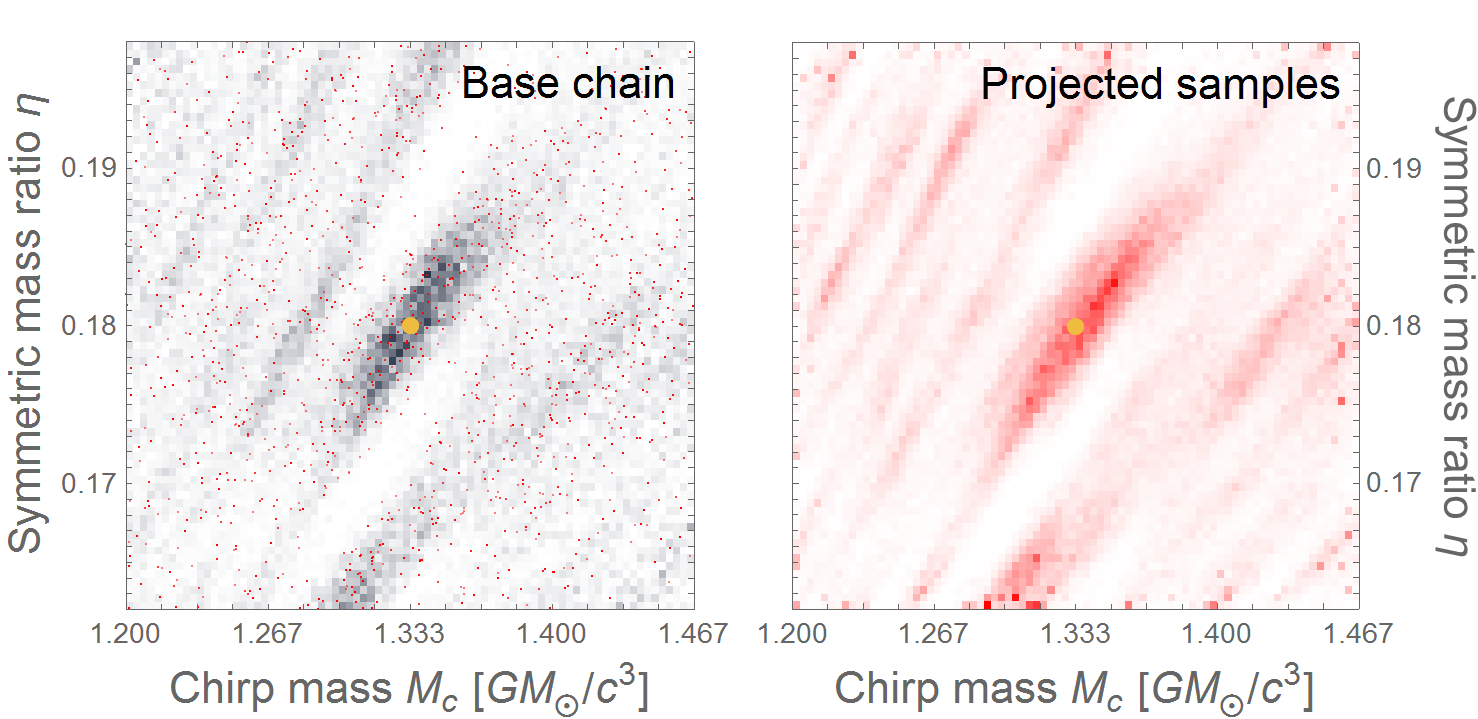}
\caption{Using the derivative-based sampling method of \cite{2018arXiv181105494C}, a ``base chain'' (left; red points) is formed from the first 5,000 samples of the $10^5$-sample chain in Fig.\,\ref{fig:fisher}. Each base point seeds a ``mini-distribution'' of 20 points that are projected onto the tangent bundle of the signal manifold, and weighted accordingly in an approximate density histogram (right).}
\label{fig:tangentbundle}
\end{figure}

Another promising application is derivative-based sampling, which has hitherto been underutilized in GW parameter estimation due to the dearth of tractable waveform/likelihood derivatives. Most such techniques are Markov-chain Monte Carlo algorithms with gradient-informed chain dynamics (e.g., the Metropolis-adjusted Langevin algorithm \cite{doi:10.1111/1467-9868.00123,doi:10.1111/j.1467-9868.2010.00765.x}, Hamiltonian Monte Carlo \cite{DUANE1987216}, and a family of stochastic-gradient variants \cite{Ma:2015:CRS:2969442.2969566}). They improve convergence near the maximum-likelihood point, but appear to be of less benefit in the more difficult global-search problem (due to many suppressed stationary points in the tail of a typical GW likelihood, which are present even at high SNR and can be found by mapping out the gradient field within our framework). A different type of derivative-based method for local sampling and density estimation exploits the constrained-Gaussian form of a target density such as Eq.\,\eqref{eq:redlike} to produce approximate samples efficiently \cite{2018arXiv181105494C}. Applying this scheme to the above example gives the estimated histogram in Fig.\,\ref{fig:tangentbundle}, with order-of-magnitude computational savings.

\paragraph{Conclusion.}
\label{sec:conclusion}

We submit that ANNs are powerful tools for the high-dimensional interpolation of reduced-basis projection coefficients $\alpha(\theta)$, as necessary for the application of ROM to GW data analysis. Our approach is suitable for any waveform model whose signal space can be represented by a compact ($d\sim10^2$) reduced basis; more extensive parameter domains may be dealt with piecewise. The ANNs provide fast, reliable derivatives that enable new techniques in GW parameter estimation. Another intriguing prospect is the possibility of inverting the ANN into a map from signal space to parameter space, in effect using ROM coefficients (obtained by projecting detector data onto the reduced basis) as natural machine-learning features.
Such inverse ANNs could be trained on noisy data to provide quick maximum-likelihood estimates, supplemented by Fisher matrices from the forward map; their construction is left for future work.

\paragraph{Acknowledgements.}

We thank Natalia Korsakova and Michael Katz for helpful conversations, and we acknowledge feedback from fellow participants in the 2018 LISA workshop at the Keck Institute for Space Studies. This work was supported by the JPL Research and Technology Development program, and was carried out at JPL, California Institute of Technology, under a contract with the National Aeronautics and Space Administration. \copyright\,2019 California Institute of Technology. U.S. Government sponsorship acknowledged.

\bibliographystyle{unsrt}
\bibliography{references}

\end{document}